\documentstyle[12pt,epsf]{article}
\input epsf

\renewcommand{\thefootnote}{\fnsymbol{footnote}}
\newcommand{\nn}{\nonumber\\[6pt]}
\newcommand{\lb}[1]{\label{#1}}
\newcommand{\p}[1]{(\ref{#1})}
\newcommand{\beq}{\begin{equation}}
\newcommand{\eeq}{\end{equation}}
\newcommand{\be}{\begin{eqnarray}}
\newcommand{\ee}{\end{eqnarray}}

\begin{document}
\thispagestyle{empty}

\begin{flushright}
SUBATECH--2004--01\\
JINR--E2--2004--11 \\
\end{flushright}

\vspace{1.2cm}

\begin{center}

{\Large\bf  Symplectic sigma models in superspace }

\vspace{0.8cm}

{\Large  E.A. Ivanov} \\
\vspace{0.4cm}

{\it Bogoliubov  Laboratory of Theoretical Physics, JINR, 141980 Dubna,
Russia}\\
\vspace{0.2cm}

{\tt eivanov@thsun1.jinr.ru}
\vspace{0.5cm}

and

\vspace{0.5cm}

{\Large A.V. Smilga} \\

\vspace{0.4cm}

{\it SUBATECH, Universit\'e de
Nantes,  4 rue Alfred Kastler, BP 20722, Nantes  44307, France
\footnote{On leave of absence from ITEP, Moscow, Russia.}}
\vspace{0.2cm}

{\tt Andrei.Smilga@math.univ-nantes.fr}

\end{center}

\vspace{1.1cm}

\begin{abstract}
\noindent We discuss a special ``symplectic'' class of ${\cal N} = 4$
supersymmetric sigma models in $(0+1)$ dimension with $5r$ bosonic and $4r$
complex fermionic degrees of freedom. These models can be described off shell
by  ${\cal N} = 2$ superfields (so that only half of supersymmetries
are manifest) and also by  ${\cal N} = 4$ superfields in the framework
of the harmonic superspace approach. Using the latter, we derive the general form of the relevant
bosonic target metric.
\end{abstract}


\vfill

\newpage

\renewcommand{\thefootnote}{\arabic{footnote}}
\setcounter{footnote}0
\setcounter{page}{1}

\section{Introduction}

Three classes of supersymmetric sigma models (in dimensions $d\leq 4$)
are widely known:
generic, K\"ahler, and hyper--K\"ahler. In all these models,
the fermions are vectors in  tangent space so that the number of dynamical
(on-shell)
 fermionic degrees of freedom coincides with the dimension of the bosonic
 target manifold.
The component action of all such models has the same generic form
 and the presence of extra supersymmetries is due to some special properties
of  K\"ahler resp.  hyper--K\"ahler manifolds. A theorem proven in \cite{Freedman}
claims that no other supersymmetric sigma model can be constructed.
This is true under two assumptions: {\it (i)} the kinetic part of the Lagrangian
depends only on the metric in the standard way $\sim g_{ij}(\phi) \partial_\mu
\phi^i \partial_\mu \phi^j$; {\it (ii)} the fields $\phi^i$ depend on time and at least
one spatial coordinate.

In two dimensions, the first condition can be relaxed by adding the torsion term
$\sim \epsilon_{\mu\nu} h_{ij}(\phi) \partial_\mu
\phi^i \partial_\nu \phi^j$.
If such a term is allowed, one can construct a twisted supersymmetric
sigma model, which enjoys  ${\cal N} = 4$ supersymmetry
\footnote{In our convention here ${\cal N}$ always counts the number of {\it complex}
supercharges; then ${\cal N}=4$ in $d=2$ means two complex supercharges both
in the left and right
light-cone sectors.}
even though the target space is not hyper--K\"ahler (neither it is K\"ahler)
\cite{GHR}.\footnote{The relevant geometry is sometimes called ``hyper-K\"ahler
with torsion''
(HKT)\cite{HKT}.} If the second condition is not satisfied and we are dealing
not with field theory, but with quantum mechanics, even a larger variety of
supersymmetric sigma models can be constructed. A special interesting subclass of
such $(0+1)$ sigma models is formed by those which cannot be directly reproduced
by dimensional reduction from the higher $d$ ones.

The simplest model of this type is defined on a conformally flat 3--dimen- \break sional
manifold with fermions and bosons belonging, respectively, to the fundamental and adjoint
representations of $SO(3)$ $\sim $ $USp(2)$. The Lagrangian in components was
constructed in \cite{SQED} and the superfield description was given in \cite{my} (see also \cite{bp}).

Let us dwell on the latter in some more details. We introduce the real superfield
$V_k(t,\theta_\alpha, \bar\theta^\alpha)$,\ $(k =1,2,3);\
\bar \theta^\alpha = (\theta_\alpha)^\dagger$, or, in spinor notation,
  \be
  \lb{spin-vect}
V_{\alpha\beta} \ =\ i(\sigma_k)_\alpha^{\ \, \gamma} \epsilon_{\beta\gamma} V_k\,,
\quad
 V_{\alpha\beta} = V_{\beta\alpha}\, ,
  \ee
where $\epsilon_{\beta\gamma} = -\epsilon_{\gamma\beta}\,, \;
\epsilon_{\alpha\beta}\epsilon^{\beta\gamma} = \delta^\gamma_\alpha$ and
$\epsilon_{12} = \epsilon^{21} = 1$.
Let us impose the constraint
  \be
\label{albetgam}
D_{(\alpha} V_{\beta\gamma)} \ =\ 0\,, \quad \bar D_{(\alpha}V_{\beta\gamma)} = 0
  \ee
where
\be
D_\alpha = \frac{\partial}{\partial \theta^\alpha} + i\bar\theta_\alpha
\frac{\partial}{\partial t}\,, \;\;
\bar D_\alpha = \frac{\partial}{\partial \bar\theta^\alpha} - i\theta_\alpha
 \frac{\partial}{\partial t}
\ee
are the covariant derivatives, $\bar D_\alpha = (D^\alpha)^\dagger$,
and the doublet $SU(2)$ indices are raised
and lowered with the help of the invariant tensors $\epsilon^{\alpha\beta},
\epsilon_{\alpha\beta}\,$.
The constrained superfield $V_k$ is expressed in terms of components as follows:
\be
\label{Vcomp}
V_k &=& A_k + \bar\psi \sigma_k \theta + \bar\theta
\sigma_k \psi + \epsilon_{kjp} \dot{A}_j \bar\theta
\sigma_p \theta + D \bar\theta \sigma_k \theta \nn
&& +\, i(\bar\theta \sigma_k \dot{\psi} -
\dot{\bar\psi}
\sigma_k \theta )\bar\theta \theta + \frac {\ddot{A}_k}4
\theta^2 \bar\theta^2
 \ee
 ($\theta^2 = \theta^\alpha \theta_\alpha, \ \bar \theta^2 = \bar \theta_\alpha \bar \theta^\alpha$,
 $\bar\theta\theta = \bar\theta^\alpha\theta_\alpha\,$).
We have, as promised, 3 bosonic ($A_k$) and 2 complex fermionic ($\psi_\alpha$)
 variables.
 $D$ is the auxiliary field. By construction, the action
 \be
\label{act3}
S \ =\ \int dt \int d^2 \theta d^2 \bar\theta\  F(V_k)
 \ee
enjoys off-shell ${\cal N} = 2$ supersymmetry.

The action (\ref{act3}) is genetically related
to the action of $4D$  ${\cal N} =1$ supersymmetric gauge models. The superfield
$V_{\alpha\beta}$ can be represented as
  \be
  \label{VkviaC}
V_{\alpha\beta} =
( D_\alpha \bar D_\beta + D_\beta \bar D_\alpha )C \
  \ee
where $C$ is the standard vector superfield prepotential reduced to $(0+1)$ dimensions.
The standard action of the dimensionally reduced supersymmetric photodynamic
$\propto \int dt \int d^2\theta\  W_\alpha^2$ with
$W_\alpha \propto (\bar D)^2D_\alpha C$ can be cast in the form (\ref{act3})
with $F_0(V) = V_k^2$. The point is that in the quantum mechanical limit $V_k$ is {
\it gauge invariant}, and now {\it any} function $F(V_k)$ is allowed as the invariant
Lagrangian. Actually, $V_k$ is just the spatial part of the initial $4D$ vector $U(1)$
superconnection. In the case of general Lagrangian function $F(V_k)$, it is impossible to rewrite the
invariant action only in terms of $W_\alpha$, so this action cannot be recovered
from a $4D$ supersymmetric gauge-invariant action by the direct dimensional reduction
$3+1   \rightarrow 0+1$.
The kinship to gauge models displays itself in the fact that the model (\ref{act3})
was first derived in \cite{SQED} (at the component level) as the effective action for supersymmetric QED in
small
spatial volume.

An obvious generalization of the model (\ref{act3}) is
 \be
\label{act3many}
 S \ =\ \int dt \int d^2 \theta d^2 \bar\theta\  F(V_k^A),\ \ \ \ \ \ \ \ \
 A = 1,\ldots,r\,.
 \ee
It involves $r$ different fermion variables $\psi_\alpha^A$ belonging to the
fundamental representation of $USp(2)$ and $r$ bosonic variables belonging to the
 adjoint representation. Such a model (with $F$ of some special form) represents the effective action of a
{ non--Abelian}
gauge theory in small volume, with $A$ running over Cartan generators and $r$ being
the rank
of the gauge group \cite{BO}. A remarkable fact is that the model (\ref{act3many})
describes
also the dynamics of slowly moving maximally charged Reissner--Nordstr\"om
black holes \cite{Strom}.
\footnote{This duality is presumably a reflection of
the general ``Gauge theories/Strings'' correspondence \cite{ads}.} The correct
expression of the
target bosonic metric in  the generic ``multiflavor'' case can be found in
\cite{Strom,IL}.

Another class of models was first introduced in ref. \cite{DE}
(for the accurate expression of the Lagrangian in components,
see ref.\cite{N2}). In its simplest variant,
the model is defined on a 5--dimensional manifold
$v_I \equiv (v_k, \phi, \bar \phi)$ ($k=1,2,3$) and involves
4 fermion degrees of freedom (belonging to the fundamental representation of
$SO(5) \sim  USp(4)$).
\footnote{As explained in \cite{N2}, these models are genetically related to
 $4D$ ${\cal N} =2$ and $6D$ ${\cal N} =1$ gauge theories: they describe
low--energy dynamics of such theories in small box. The $SO(5)$ covariance
of the  $(0+1)$ dimensional models is related to spatial rotational symmetry of
$6D$ theory. }
The metric of the model is conformally-flat like in the model
discussed earlier and it is derived from a prepotential
${\cal K}(v_k, \phi, \bar \phi)$ as
  \be
\label{metr5}
 g_{IJ} \ =\ 2 \delta_{IJ} \frac {\partial^2 {\cal K}}{\partial v_k \partial v_k}
\equiv \delta_{IJ}\,h\,.
  \ee
The function ${\cal K}$ is subject to
the constraint (5-dimensional Laplace equation)
\be
\Delta^{(5)} {\cal K} = \frac {\partial^2 {\cal K}}{\partial v_k \partial v_k} +
2\frac {\partial^2 {\cal K}}{\partial \phi \partial \bar\phi} = 0\,.\label{00}
\ee
In ref.
\cite{DE}, the Lagrangian was written as a function of the ${\cal N} = 2$
superfields $V_k$ discussed
above and standard chiral superfields $\Phi, \bar\Phi$. The off-shell action
 has the form
  \be
 \label{act5}
S = \int dt \int d^2 \theta d^2 \bar\theta \ {\cal K} (V_k, \bar\Phi, \Phi)\, .
 \ee
It is invariant under {\it additional} ${\cal N}=2$ supersymmetry transformations
 \be
\label{trans}
\delta \bar\Phi &=& \frac {2i}3 \epsilon^\alpha (\sigma_k)_\alpha^{\ \beta}
 D_\beta V_k\,, \nonumber \\
\delta \Phi &=&  \frac {2i}3 \bar \epsilon_\alpha (\sigma_k)_\beta^{\ \alpha}\bar D^\beta V_k\,,
\nonumber \\
\delta V_k &=& -i\epsilon^\alpha (\sigma_k)_\alpha^{\ \beta} D_\beta \Phi -
i \bar \epsilon_\alpha (\sigma_k)_\beta^{\ \alpha} \bar D^\beta \bar \Phi\,,
 \ee
provided that
  \be
\label{harm5}
  \frac {\partial^2{\cal K}}{\partial V_k^2} +
2 \frac {\partial^2{\cal K}}{\partial \bar\Phi \partial \Phi}
 \ =\ 0\ .
  \ee
When restricted to the bosonic target
manifold $v_k = V_k|$, $\phi = \Phi|$, $\bar\phi = \Phi|$ (hereafter,
``$X|$''  stands for the lowest component of the superfield $X$), eq. \p{harm5}
yields just the 5-dimensional Laplace
equation \p{00}. In the next section we explain in some more details how the
constraint (\ref{harm5})
can be derived. Then we consider the generalization of (\ref{act5}) to the
``multiflavor''
case, derive a set of constraints for the prepotential ${\cal K}$ and present
their
solution. In Sect. 3 we present the description of these models in terms of
${\cal N} = 4$
harmonic superfields and discuss the relationship between the ${\cal N} = 2$
and ${\cal N} = 4$
approaches.

In what follows, the $d=1$ sigma models with $3r$- and
$5r$-dimensional bosonic target manifolds we have defined above
will be referred to as {\it symplectic} sigma models, respectively of
the first and second types. This nomenclature emphasizes the property that their bosonic and fermionic fields
belong to the adjoint and fundamental representations of the symplectic
groups $USp(2) \sim SO(3)$ and $USp(4) \sim SO(5)$.\footnote{To avoid a misunderstanding, let us
point out that these symmetries can be explicitly broken in the sigma model actions
and so are not isometries of the general bosonic target metric.}

\section{${\cal N} = 2$ description}
\setcounter{equation}0

Consider the variation of (\ref{act5}) under the transformation (\ref{trans}). The term
involving $\epsilon^\alpha$ is proportional to
     \be
\int dt \int d^4\theta \ (\sigma_k)_\alpha^{\ \beta} \left[ \frac 23 \frac {\partial
{\cal K}}{\partial \bar \Phi} D_\beta V_k - \frac {\partial {\cal K}}{\partial V_k}
D_\beta \Phi
\right].
  \ee
For the integral to vanish, the integrand has to be a total (covariant) derivative,
i.e.
\be
\label{uslovie}
(\sigma_k)_\alpha^{\ \beta} \left[ \frac 23 \frac {\partial
{\cal K}}{\partial \bar \Phi} D_\beta V_k - \frac {\partial {\cal K}}{\partial V_k}
D_\beta \Phi
\right] = D_\alpha R + (\sigma_k)_\alpha^{\ \beta} D_\beta L_k
 \ee
where $R$ and $L_k$ are some complex functions of the involved superfields.
Let us write
 \be
D_\alpha R &=& \frac {\partial R}{\partial \Phi} D_\alpha \Phi +
 \frac {\partial R}{\partial V_k} D_\alpha V_k\,, \nonumber \\
 D_\beta L_k &=& \frac {\partial L_k}{\partial \Phi} D_\beta \Phi +
 \frac {\partial L_k}{\partial V_p} D_\beta V_p \label{DRL}
  \ee
where we used the chirality condition $D_\alpha \bar\Phi =0$. Next we need to take
into
account the constraint (\ref{albetgam}) which leaves only the irreducible
$\bf \frac 12$ part in the superfield
$D_\alpha V_k$ (transforming as the tensor product
$\bf \frac 12 \otimes 1 = \frac 12 \oplus \frac 32$),
    \be
D_\alpha V_k = (\sigma_k)_\alpha^\beta\chi_\beta\,.
   \ee
Then, comparing the coefficients of the independent structures in both sides of
\p{DRL}, we
end up with the set of linear differential equations
  \be
\label{integriruemost}
2\frac {\partial {\cal K}}{\partial \bar\Phi} - \frac {\partial L_k}{\partial V_k}
&=& 0\,,
\quad (\mbox{a})
\nonumber \\
i\epsilon_{kpq} \frac {\partial L_k}{\partial V^p} + \frac {\partial R}{\partial V_q}
&=& 0\,, \quad (\mbox{b})\nonumber \\
\frac {\partial {\cal K}}{\partial V_k} + \frac {\partial L_k}{\partial \Phi} &=& 0\,,
\quad (\mbox{c})
\nonumber \\
\frac {\partial R}{\partial \Phi} &=& 0\,. \quad (\mbox{d})
  \ee

Differentiating eq. (\ref{integriruemost}a) with respect to $\Phi$ and eq.
(\ref{integriruemost}c)
 with respect to $V_k$
and summing up the obtained relations, we
arrive at eq. (\ref{harm5}) as an integrability condition for the set
\p{integriruemost}.
The rest of equations imposes some constraints on the unknowns $R$ and $L_k$
the precise generic form of
which is of no interest for our purposes. Nevertheless, we should show that eq.
 (\ref{harm5}) is not only
necessary, but also sufficient condition for the system (\ref{integriruemost})
to have a nontrivial solution.
It is not difficult to check that
    \be
R = 0 ,\ \ \ \ \ \ \ \ \ \  L_k = \frac {\partial Q} {\partial V_k} \label{solut1}
\ee
supplies a particular solution provided $Q$ is a 5-dimensional harmonic function
such that ${\cal K} =
- \partial Q/\partial \Phi$. For the Green's function of the Laplace equation
 \be
\label{Green5}
{\cal K} \ =\ \frac 1{(2\bar\Phi \Phi + V_k^2)^{3/2}}\ ,
  \ee
we have
 \be
Q \ =\ \frac 1{\bar\Phi (2\bar\Phi \Phi + V_k^2)^{1/2}} \ \ \ \ \ {\rm and} \ \ \ \ \
L_k \ =\ -\frac {V_k}{\bar\Phi (2\bar\Phi \Phi + V_k^2)^{3/2}}
  \ee
(it is assumed that the range of the allowed values of the involved superfields does
not
include the singularity point $\Phi = \bar\Phi = V_k =0$).
An arbitrary harmonic function $Q$ can be represented as a linear superposition of
the Green's functions (\ref{Green5}).

By  the same token as for the symplectic $\sigma$ model of the first kind discussed
above,
the Lagrangian (\ref{act5}) can be generalized to the ``multiflavor'' case:
  \be
 \label{act5many}
S \ =\ \int dt  \int d^2 \theta d^2 \bar\theta \ {\cal K} (V_k^A, \bar\Phi^A, \Phi^A)
 \ ,
\ \ \ \ \ \ \ \ A = 1,\ldots, r \ .
 \ee
Let us find the constraints to be imposed on ${\cal K}$ for the action
(\ref{act5many}) to be
${\cal N} = 4$ supersymmetric. Proceeding along the same lines as before, we
arrive at the
system
  \be
 \label{integrA}
 2\frac {\partial {\cal K}}{\partial \bar\Phi^A} - \frac {\partial L_k}{\partial
V_k^A} &=& 0\,, \quad (\mbox{a})
  \nonumber \\
  i\epsilon_{kpq} \frac {\partial L_k}{\partial V_p^A} + \frac {\partial R}{\partial
V_q^A}
   &=& 0\,, \quad (\mbox{b})\nonumber \\
  \frac {\partial {\cal K}}{\partial V_k^A} + \frac {\partial L_k}{\partial \Phi^A}
 &=& 0\,, \quad (\mbox{c})
   \nonumber \\
   \frac {\partial R}{\partial \Phi^A} &=& 0\,.  \quad (\mbox{d})
     \ee

Comparing eqs. (\ref{integrA}a) and (\ref{integrA}c), we obtain a kind of the
generalized harmonicity condition
  \be
 \label{harmAB}
\left(2 \frac {\partial^2}{\partial \bar\Phi^A \partial \Phi^B} + \frac
{\partial^2}{\partial
  V_k^A \partial V_k^B} \right) {\cal K} \ =\ 0\ ,
   \ee
as the integrability condition for these equations. Eq. \p{harmAB} actually amounts
to two independent
conditions which are obtained by symmetrizing and antisymmetrizing indices $A,B$.
The antisymmetric
condition reads
\be
\frac {\partial^2 {\cal K}}{\partial \Phi^{[A} \partial \bar\Phi^{B]}}
\ =\ 0\ .\label{antisymm}
\ee
This is not the end of the story. Differentiating eq. (\ref{integrA}c) with
respect to the argument $\Phi^B$ and antisymmetrizing,
we obtain the condition
 \be
\label{VFiAB}
\frac {\partial^2 {\cal K}}{\partial V_k{}^{[A} \partial \Phi^{B]}} \ =\ 0\ .
 \ee
Its complex conjugate is
 \be
\label{VFibarAB}
\frac {\partial^2 {\cal K}}{\partial V_k{}^{[A} \partial \bar\Phi^{B]}} \ =\ 0\ .
 \ee
Finally, differentiating eq. (\ref{integrA}b) with respect to  $\Phi^B$ and eq.
(\ref{integrA}c) with respect to $V_p^B$, and
then making use of eq. (\ref{integrA}d), we obtain one more integrability
condition
 \be
\label{Vkp}
\frac {\partial^2 {\cal K}}{\partial V_k^{[A} \partial V_p^{B]}} \ =\ 0\ .
   \ee
The concise 5--dimensional form of all the conditions derived is
   \be
\label{harm5r}
\frac {\partial^2 {\cal K}}{\partial V_I^{A} \partial V_I^{B}}
\ =\ 0\ , \ \ \ \ \ \ \ \ \ \
 \frac {\partial^2 {\cal K}}{\partial V_I^{[A} \partial V_J^{B]}}
\ =\ 0\,,
    \ee
where $V^A_I \equiv (V^A_k, \mbox{Re}\Phi^A, \mbox{Im}\Phi^A)$.
The first condition is just the symmetric part of \p{harmAB},
while the second one encompasses \p{antisymm} - \p{Vkp}.

  An example of the function ${\cal K}$ satisfying the constraints (\ref{harm5r})
is a harmonic function $h(V^1_I)$ (no dependence on $V^{2}_I,\ldots,V^r_I$). A more
general solution is $h(p_A V^A_I + X_I)$ with arbitrary $X_I$ and $p_A$ (at
first glance, the constant shift $X_I$
seems to be redundant; the reason why it was included is explained a few lines lower).
 Still more
generally, one can write ${\cal K}$ as a sum of several terms
   \be
\label{solK}
{\cal K} \ =\ \sum_n \ h^{(n)} \left(p_A^{(n)} V^A_I  + X_I^{(n)} \right).
 \ee
Eq. (\ref{solK}) can be further generalized by upgrading the sum $\sum_n$ to an
integral.
Our {\it conjecture} is that in this way one can obtain a {\it general} solution
for the set
(\ref{harm5r}). An important remark is in order. Mathematicians like to solve the
Laplace
equation and other elliptic equations in a finite region, imposing Dirichlet or
Neumann, or mixed conditions
at the boundary. For physical applications, we rather  need to solve eqs.(\ref{harm5r})
 in
the whole $R^{5r}$, allowing certain singularities inside. The Green's function
(\ref{Green5})
is such a solution for the Laplace equation with singularity at origin. The harmonic
functions
entering the sum in eq.(\ref{solK}) should be also thought of as such Green's function
 which are singular
at $5(r-1)$ -- dimensional hyperplanes $p_A^{(n)} V^A_I  + X_I^{(n)} = 0$. As soon
as {\it some} region
in $R^{5r}$ is free from singularities, the solution makes sense.

  In fact, the model (\ref{act5many}) was derived as the effective Lagrangian of 4D
${\cal N} = 2$
supersymmetric non--Abelian gauge theory in a small box \cite{N2}. In that case, the
function
${\cal K}$ was shown to have the form (\ref{solK}) with $X_I^{(n)} = 0$ and $p_A$
having the meaning
of the roots in the Lie algebra of the corresponding gauge group.

 When ${\cal K}$ is expressed as a sum over Green's  functions (\ref{Green5}) of the
argument $p_A^{(n)} V^A_I  + X_I^{(n)}$, the solution to the system of linear
equations (\ref{integrA})
for $R,L_k$ can be easily found as a sum of the corresponding expressions for
$L_k$ (the function $R$ can be
set equal to zero).
 The metric that follows from eq.(\ref{act5many}) has the form
   \be
\label{metr5AB}
 g_{IJ}^{AB} \ =\ \delta_{IJ} h^{AB} = 2 \delta_{IJ} \frac {\partial^2
{\cal K}}{\partial v^A_k \partial v_k^B}\ .
  \ee

\section{${\cal N}=4$ description in harmonic superspace}
\setcounter{equation}0

Like the $3r$-dimensional symplectic model (\ref{act3many}) is related to Abelian
${\cal N} = 1,$ $4D$
supersymmetric photodynamics , the model (\ref{act5many}) is related to ${\cal N} = 2,$
 $4D$
Abelian supersymmetric photodynamics. Using the technique of harmonic superspace (HSS)
 \cite{GIKOS,kniga},
the action for this model can be written in such a way that all 4 complex
supersymmetries are manifest.

\subsection{Overview of ${\cal N}=2$ photodynamics in HSS}
We start by recalling some salient features of the HSS description of ${\cal N}=2$
 photodynamics.

The basic building block is the harmonic analytic superfield
$$V^{++}(x^m_A,
\theta_\alpha^+,\bar\theta^{+}_{\dot{\alpha}},
u^\pm_i) \equiv V^{++}(\zeta^M, u)$$
where $u^\pm_i$ are spherical spinorial
harmonics parametrizing the $R$-symmetry coset $S^2 \sim SU(2)_R/U(1)_R$,
$u^-_i = (u^{+i})^\dagger \Longrightarrow  \ u^{+i}u^-_i =1$ ($i=1,2$) and $\theta^+_\alpha =
\theta_{\alpha}^{ i} u^{+}_i,
\bar\theta^+_{\dot\alpha} = \bar\theta_{\dot\alpha}^{ i} u^{+}_i,$ with
$\bar \theta_{\dot{\alpha}}^i = (\theta_{i\alpha})^\dagger$. The superfield
$V^{++}$
does not depend on another half of Grassmann coordinates, i.e. on $\theta^-_\alpha =
\theta_{\alpha}^{ i} u^{-}_i,
\bar\theta^-_{\dot\alpha} = \bar\theta_{\dot\alpha}^{ i} u^{-}_i$, and in this sense
it is
Grassmann-analytic \cite{GIO} like a chiral superfield.
The ``analytic basis''
coordinate $x^m_A$
is related to $x^m$ (the ``central basis'' coordinate) by a $\theta$-dependent shift,
$$x^m_A \ =\ x^m - i\theta^+ \sigma^m \bar \theta^- -
i\theta^- \sigma^m \bar \theta^+ \ ,$$
similar to left and right coordinates in ${\cal N} = 1$ chiral superspace.
The harmonic analyticity is covariantly expressed
as the Grassmann Cauchi-Riemann conditions
\be
D^{+}_\alpha\,V^{++} = 0\,, \quad \bar D^{+}_{\dot\alpha}V^{++} = 0 \ , \lb{CR}
\ee
where
\be
D_\alpha^+ = u^+_i D^i_\alpha\,, \quad \bar D_{\dot\alpha}^+ = u^+_i
\bar D^i_{\dot\alpha}\,,\lb{Proj}
\ee
and $D^i_\alpha$ and $\bar D^i_{\dot\alpha}$ are standard spinor covariant derivatives
with respect to Grassmann coordinates
in the central basis $(x^m, \theta_{i\alpha}, \bar\theta^i_{\dot\alpha},
u^\pm_k) \equiv (X^M, u)\,$.\footnote{We basically follow conventions and notation of
ref. \cite{kniga}.} The
harmonic projections \p{Proj} satisfy the following anticommutation relations
\be
\{D^-_\alpha, \bar D^+_{\dot\alpha} \} = -\{D^+_\alpha, \bar D^-_{\dot\alpha} \} =
2i(\sigma^m)_{\alpha\dot\alpha}P_m\,,
\lb{AComm}
\ee
all other (anti)commutators being zero. An important difference of the {\it harmonic}
Grassmann analyticity
from the chirality is that $V^{++}$ (and any other analytic superfield with
the even $U(1)$ charge) can be chosen real with respect to a special
involution which preserves the set
of analytic coordinates and acts as the standard complex conjugation on any component
field.

The fields
in the $\theta^+, \bar\theta^+$-expansion
of $V^{++}$ are expanded in a harmonic series over $u^\pm_i$. Now, ``$++$'' upstairs
means that only
the terms of $U(1)$ charge +2
(i.e. the terms $\propto (u^+)^2, \ (u^+)^3 u^-$, etc) are kept in the harmonic
expansion of the
lowest component $v^{++} = V^{++}|$.
The rest of component fields carry $U(1)$ charge $q$ ranging from $+1$ to $-2$, in
accord with the obvious rule that
the sum of $q$ and the charge of the relevant $\theta$-monomial is always $+2$.
The corresponding harmonic expansions go over the $u^\pm_i$-monomials having the
charge $q\,$.

Because of harmonic expansions,
$V^{++}$ encompasses an infinite number of field components. Fortunately, most of
them are pure gauge degrees of freedom
thanks to the invariance under the gauge transformation
\be
V^{++} \;\;\Rightarrow \;\; V^{++}\,{}' = V^{++} + D^{++}\Lambda\,. \lb{gauge4}
\ee
Here, the gauge superparameter $\Lambda$ is an unconstrained harmonic analytic
superfield involving an infinite
number of ordinary gauge parameters, and $D^{++}$ is the harmonic derivative which
preserves harmonic Grassmann analyticity
\be
D^{++} &=& \partial^{++} - 2i\theta^+\sigma^m\bar\theta^+\partial_m
+ \theta_\alpha^+ \frac {\partial}{ \partial \theta_\alpha^-}
+ \bar \theta_{\dot{\alpha}}^+  \frac {\partial}{ \partial \bar\theta_{\dot{\alpha}}^-} \nonumber \\
\partial^{++}
&=&  u^{+i}\frac{\partial}{\partial u^{-i}}\,.
\ee
This was written in the analytical basis $(x^m_A, \theta^\pm_\alpha,
\bar \theta_{\dot{\alpha}}^\pm, u)$.
In the central basis, $D^{++} $ coincides with $\partial^{++}$ . One can equally well introduce
the derivative $D^{--}$ coinciding with
\be
\partial^{--} \ =\ u^{-i} \frac {\partial}{\partial u^{+i}}
 \ee
in the central basis. Irrespective of the choice of the basis, the following commutation
relations hold:
 \be
&&[ D^{\pm\pm}, D^\pm_\alpha ] = 0\,, \quad  [D^{\pm\pm}, D^\mp_\alpha] = D^\pm_\alpha\,,
\quad  {\rm and\  c.c.} \label{com++-} \\
&& [  D^{++}, D^{--} ] =  D^0 \nonumber
\ee
where $D^0$ is the operator counting external $U(1)$ charges of functions on $S^2$,
$D^0\,\Psi^q(u) = q\,\Psi^q (u)$. In the central basis it reads
$$
D^0 = u^{+i} \frac {\partial}{\partial u^{+i}} -
 u^{-i} \frac {\partial}{\partial u^{-i}}\,.
$$
An important property is
  \be
\label{Du}
 \int du  \, d^4x D^{++}  F(u,x) \ =\ \int du d^4x \,D^{--} F(u,x)\ =\ 0\ ,
  \ee
which allows one to integrate by parts with respect to $D^{++}$ and
$D^{--}$.

The gauge freedom \p{gauge4} can be
used to gauge away all fields in $V^{++}$ except for $(8 + 8)$ fields of the off-shell
vector ${\cal N}=2$
multiplet \cite{GIKOS,kniga}. In such a gauge (${\cal N} = 2$ version of
the Wess-Zumino gauge), $V^{++}$ looks as
\be
\label{V++}
 V^{++}_{WZ}(\zeta^M, u)  = \sqrt{2} (\theta^+)^2\bar\phi(x) + \sqrt{2} (\bar\theta^+)^2\phi(x)
-2i \theta^+\sigma^m\bar\theta^+ A_m(x) \nn
 + 4(\bar\theta^+)^2\theta^{+\alpha}\psi^i_\alpha(x) u^-_i -
4(\theta^+)^2\bar\theta^+_{\dot\alpha}\bar\psi^{\dot\alpha i}(x)u^-_i
+ 3(\theta^+)^2(\bar\theta^+)^2 G^{ik}(x)u^-_iu^-_k \lb{WZ1}
\ee
where, for  brevity,  we suppressed the index ``$A$'' of $x^m_A$. The physical fields
are  $\phi, A_m, \psi^i_\alpha$.  The coefficients in (\ref{V++})
are chosen so that the kinetic terms for all these fields
in the action (\ref{actphot}) to be defined below are normalized in a standard
way. Now,
$G^{ik}$ is the $SU(2)_R$ triplet of auxiliary fields.

For constructing covariant superfield strength and invariant actions one needs also
the general (non-analytic)
harmonic superfield $V^{--}$ which is related to $V^{++}$ by the harmonic equation
   \be \lb{basicVV}
D^{++} V^{--} \ =\  D^{--} V^{++}\ .
  \ee
 The requirement for eq. \p{basicVV} to be gauge covariant dictates that $V^{--}$ is
transformed as
\be \lb{gauge42}
V^{--}\;\;\Rightarrow \;\; V^{--}\,{}' = V^{--} + D^{--}\Lambda
\ee
(use the last relation in (\ref{com++-}) and the fact that $ \Lambda $ has zero $q$ charge).
The action has the form
  \be
 \label{actphot}
S =   \frac 14 \int du\ d^{12}X\ V^{++} V^{--}\ .
  \ee
Its gauge invariance can be checked using \p{basicVV} and the analyticity conditions
\p{CR} for $V^{++}$ and $\Lambda$.

The action \p{actphot} involves, besides
$\int d^{12}X \equiv \int d^4x \,d^4\theta d^4\bar\theta\,$, also
the integral $\int du$ over $S^2 \sim SU(2)_R/U(1)_R$.
The integral $\int du$ can actually be explicitly done (see below),
after which \p{actphot} takes
the standard form of an integral over chiral ${\cal N}=2$ superspace. It admits three
equivalent representations
   \be
S &=& \frac 14 \int d^4x_L d^4\theta \ W^2  =
\frac 14 \int d^4 x_R d^4\bar\theta \ \bar W^2  \nn
&=&   \frac 18 \int d^4 x_L d^4\theta \ W^2  +  \frac 18 \int d^4x_R d^4\bar\theta \
\bar W^2\,. \label{W2}
  \ee
Here
\footnote{$(D^+)^2 \equiv D^{+\alpha}D^+_\alpha\,, \;(\bar D^+)^2 \equiv
\bar D^+_{\dot\alpha}\bar D^{+\dot\alpha}\,$.}
  \be
\label{Wdef}
W \ =\ - \frac 14 (\bar D^{+})^2 V^{--}\,, \quad \bar W\ =\ - \frac 14 (D^+)^2 V^{--}
   \ee
are mutually conjugated superfield strengths of Abelian ${\cal N}=2$
vector multiplet. They satisfy the chirality conditions
\be
\bar D^{\pm}_{\dot\alpha}W =0\,, \quad D^{\pm}_\alpha \bar W = 0 \lb{chirality}
\ee
and the additional irreducibility constraint
\be
(D^\pm)^2W = (\bar D^\pm)^2\bar W\,, \quad (D^+D^-)W = (\bar D^+\bar D^-)\bar W \,.
\lb{dop}
\ee

The chiral superfield strength (\ref{Wdef}) is an ${\cal N} =2$ analog of the conventional
${\cal N} =1$ field strength superfield $W_\alpha$. It is gauge invariant and
{\it does} not
depend on harmonics in the central basis. The latter follows from the property
\be
D^{++}W = D^{++}\bar W = 0\,, \lb{uindep}
\ee
which can be easily checked using \p{basicVV}, the (anti)commutation relations
and the analyticity conditions \p{CR}. Eqs. \p{uindep} also imply, via the complex conjugation, that
\be
D^{--}W = D^{--}\bar W = 0\,. \lb{uindep1}
\ee
The constraints \p{chirality} and \p{dop} actually follow from the definition \p{Wdef}
and the property
of harmonic independence of $W, \bar W$. From \p{Wdef} one immediately deduces the $+$
 and $++$
projections of \p{chirality} and \p{dop}, respectively, while the rest of these
relations is
obtained by applying $D^{--}$ to the $+$ and $++$ projections and using \p{uindep1}
together with the commutation relations.
The basic difference from  ${\cal N} =1$
superfield strength $W_\alpha$ is that (\ref{Wdef}) involves 2 rather than 3 covariant
derivatives on the gauge prepotential.
As a result, the lowest component of $W$ is a complex scalar filed rather than
a spinor one.  The explicit form of the bosonic part of $W$ in the $(0+1)$ case will be given
later in the context of  our main topic, the HSS
description of symplectic sigma models of the second kind.

Let us outline in brief how to reveal the equivalence of (\ref{actphot}) and (\ref{W2}).
The proof essentially relies upon the definition \p{Wdef} and the harmonic independence
conditions \p{uindep}, \p{uindep1}.
As a first step, one represents the integration measure in \p{actphot} as
$$
du d^{12} X = du d^4x_L d^4\theta (\bar D)^4 = du d^4 x_R d^4\bar\theta (D)^4\,,
$$
$$(D)^4 \equiv {1\over 16}(D^+)^2(D^-)^2\,,
\quad (\bar D)^4 \equiv {1\over 16}(\bar D^+)^2(\bar D^-)^2\,,\lb{split}
$$
and then act by $(\bar D)^4$ or $(D)^4$ on the integrand, in order to end up
with integrals over chiral subspaces. The representation \p{W2} is obtained
after a multiple use of analyticity of $V^{++}$, the conditions
\p{uindep}, \p{uindep1} and the commutation relations (\ref{AComm}), (\ref{com++-})
together with their corollary $[D^{\mp\mp}, (D^\pm)^2] = 2D^{\mp\alpha}D_{\pm\alpha}$.
In the process of this calculation, one should integrate by parts with respect to
harmonic derivatives using (\ref{Du}).

The point important for the following is that one can generalize the free action
 (\ref{actphot}) and write \cite{Zupn}
  \be
\label{actSW1}
S = \frac 14 \int du d^{12}X \,\left[ V^{++} V^{--}\, F(\bar W, W)\right],
\quad F|_{W=\bar W =0} = 1
\ee
where $F$ is some function of superfield strengths the form of which is not fixed
 by ${\cal N}=2$ supersymmetry alone.
However, the requirement of invariance under the gauge transformations \p{gauge4},
\p{gauge42} forces $F$ to satisfy the
stringent constraint \cite{Zupn}
\be \lb{Zupeq}
\frac{\partial^2 F}{\partial W \partial \bar W} = 0 \;\; \Rightarrow \;\; F(W,
\bar W) = f(W) + \bar f(\bar W)
\ee
where $f(W)$ is an arbitrary holomorphic function. After substituting this into
\p{actSW1} and performing the same
manipulations which lead from \p{actphot} to \p{W2}, the action \p{actSW1} takes
the form
  \be
  \label{actSW}
S &=&
\frac 14 \int du d^4x d^8\theta \;V^{++} V^{--} \left[f(W) +
\bar f(\bar W) \right] \nonumber \\
&=& \frac 14 \int d^4 x_L d^4\theta \ W^2\,f(W)  +  \frac 14 \int d^4x_R d^4\bar\theta \
\bar W^2\,\bar f(\bar W)\,.
  \ee
The famous Seiberg--Witten effective action \cite{SW} can be presented in this form.
\footnote{For the first time the action of the type \p{actSW} (in its second, harmonic-independent form)
appeared in \cite{Gates}.}

Gauge invariance of (\ref{actSW}) is manifest.
The emergence of eq. \p{Zupeq} as the necessary condition for gauge invariance of \p{actSW1} can
 be explained as follows. The variation of
$S$ under the gauge transformations \p{gauge4}, \p{gauge42} can be reduced to
\be
\delta S = - {1\over 2}\int du d^{12}X  \left[\Lambda\,D^{--}V^{++}\, F(W,
\bar W)\right]\,.
\ee
Then one represents the integration measure as
$$
du d^{12} X = du d^4 x_A (D^-)^4 (D^+)^4, \ \ \ \ \ \ \ \
(D^\pm)^4 = \frac 1{16} (D^\pm)^2 (\bar D^\pm)^2 \,,
$$
and pulls $(D^+)^4 $ through $V^{++}V^{--}$ to $F(W, \bar W)$ using the analyticity of
$V^{++}$ and
(anti)commutation relations (\ref{AComm}), (\ref{com++-}). In this process, one obtains
several terms, the most crucial of which is
$$
\propto  \int du d^4 x_A (D^-)^4 \left\{\Lambda \partial_m V^{++} [ D^{+}\sigma^m \bar
D^+ F(W, \bar W)]\right\}.
$$
It cannot be cancelled by any other term in $\delta S$, so the necessary condition of
vanishing of $\delta S$ is
\be
D^{+}_\alpha \bar D^+_{\dot\alpha} F(W, \bar W) = 0\,,
\ee
which, in virtue of the chirality conditions \p{chirality}, amounts just to \p{Zupeq}.
The remaining terms in
$\delta S$ can be shown to vanish in consequence of \p{Zupeq}.

\subsection{Manifestly ${\cal N} = 4 $ supersymmetric (0+1) sigma models}

The $(0+1)$ version of the actions (\ref{actphot}), (\ref{actSW}) can be readily
recovered by neglecting the spatial
derivatives in $V^{\pm\pm},W$ and reducing $d^4x_A  \to dt_A$. However, in the case
when there is no dependence on spatial coordinates,
the action (\ref{actSW}) can be generalized yet further, giving a manifestly
 ${\cal N}=4$ supersymmetric form
of the sigma models discussed in Sect.2.  Before explaining this point, we need to
give some necessary
formulas related to ${\cal N}=4$ HSS in $(0+1)$ dimension. Actually, the HSS formalism
for this case can be constructed
in a self-contained way in its own right, without reference to its $4D$ parent.

We shall work in the analytic basis of ${\cal N}=4, (0+1)$ HSS, parametrized by
coordinates
\be
(X^M, u^\pm_i) \equiv (t, \theta^{+}_\alpha, \bar \theta^{+\alpha}, \theta^{-}_\alpha,
\bar\theta^{-\alpha}, u^{\pm}_i)
\lb{A1}
\ee
where
$\theta^\pm_\alpha = \theta_{\alpha}^iu^{\pm}_{i}$, $\bar \theta^{\pm\alpha} =
-\bar\theta^{\alpha i}u^{\pm}_i$ and $(\theta_{\alpha i})^\dagger = \bar \theta^{\alpha i}$.
The analytic subspace closed under ${\cal N}=4$ supersymmetry is defined as
\be
(\zeta^M, u^\pm_i) \equiv (t, \theta^{+}_\alpha, \bar\theta^{+\alpha}, u^\pm_i)\,.
\lb{A2}
\ee
Both \p{A1} and \p{A2} are ``real'' under the pseudoconjugation \,
$\widetilde{}$ \,
 which is the product of the standard
complex conjugation $^\dagger$ and the Weyl reflection of $S^2 \sim SU(2)_R/U(1)_R$
\cite{GIKOS,kniga} ,
    \be
    \widetilde{t_A} = t_A\,, \;\;\widetilde{\theta^{\pm}_\alpha} =
    -\bar\theta^{\pm\alpha}\,,\;\; \widetilde{\bar\theta^{\pm\alpha}} =
     \theta^{\pm \alpha}\,, \;\;
    \widetilde{u^\pm_i} = u^{\pm i}\,.
     \ee
In what follows we omit the index $A$ on $t$, hoping that this will not be a source
of confusion.

The spinor and harmonic covariant derivatives in the analytic basis are given by
\be
&& D^+_\alpha = \frac{\partial}{\partial \theta^{-\alpha}}\,, \quad
\bar D^+_\alpha =  \frac{\partial}{\partial \bar\theta^{-\alpha}}\,, \quad
D^-_\alpha = -\frac{\partial}{\partial \theta^{+\alpha}} + 2i\bar\theta^-_\alpha
\partial_t\,, \nn
&&\bar D^-_\alpha =  -\frac{\partial}{\partial \bar\theta^{+\alpha}} +2i
\theta^-_\alpha \partial_t\,, \lb{D}
 \ee
 \be
\lb{+-com}
 \{D^+_\alpha, \bar D^-_\beta\} = -2i\epsilon_{\alpha\beta}\partial_t\,, \quad
\{D^-_\alpha, \bar D^+_\beta\}
= 2i\epsilon_{\alpha\beta}\partial_t\,, \lb{Dcomm}
  \ee
 \be
&& D^{++} = \partial^{++} -2i(\theta^{+\alpha} \bar\theta^+_\alpha )\partial_t
+ \theta^+_\alpha \frac{\partial}{\partial \theta^-_\alpha} +
\bar\theta^{+\alpha} \frac{\partial}{\partial \bar\theta^{-\alpha}}\,, \nn
&& D^{--} = \partial^{--} - 2i(\theta^{-\alpha} \bar\theta^-_\alpha) \partial_t
+ \theta^-_\alpha \frac{\partial}{\partial \theta^+_\alpha} +
\bar\theta^{-\alpha} \frac{\partial}{\partial \bar\theta^{+\alpha}}
\,. \lb{D2}
\ee
The non-vanishing commutation relations between spinor and harmonic derivatives are as
follows
\be
\lb{++-com}
[ D^{\pm\pm}, D^\mp_\alpha] = D^{\pm}_\alpha \lb{DD}
\ee
and the same for $\bar D^\pm_\alpha$.

The basic object which encompasses the irreducible off-shell set of fields of the sigma
 model
we are dealing with is the $(0+1)$ reduction of the gauge potential $V^{++}$. It is
still subject to
the gauge transformations \p{gauge4}, so we can choose the same WZ gauge as in the
 $d=4$ case, i.e. \p{WZ1}.
Since we will be interested in the bosonic target space metrics only, we can neglect
all fermions. The
bosonic target space coordinates are also singlets of the R-symmetry group $SU(2)_R$,
so we can
omit the auxiliary $SU(2)_R$ triplet $G^{(ik)}$ as well for it cannot couple to the
$SU(2)_R$
singlet sector. Thus we shall use as input the following $V^{++}_{WZ}$
\be
V^{++}_{WZ} = \sqrt{2} (\theta^+)^2 \bar\phi(t) +
\sqrt{2} (\bar\theta^+)^2\phi(t)
+ 2\theta^{+\alpha} \bar\theta^{+\beta}v_{\alpha\beta}(t)
- 2i(\theta^+ \bar\theta^+)v_0(t)\, \lb{WZ2}
\ee
with symmetric $v_{\alpha\beta}$. One can introduce $v_k$ related to $v_{\alpha\beta}$
as in eq.(\ref{spin-vect}) and observe that $v_k$ and $v_0$ are genetically
related  to the  spatial and temporal components
of the 4D  gauge vector potential $A_m$, correspondingly.
 It follows from \p{gauge4} and the explicit form of
$D^{++}$ in \p{D2}, that after dimensional reduction to (0+1) space,  the only
residual gauge freedom is
\be
v_0 \; \rightarrow \; v_0' = v_0 + \dot\lambda(t)
\ee
where $\lambda(t) = \Lambda(\zeta, u)|$. The remaining five gauge invariant fields in
\p{WZ2} are just the bosonic target space
coordinates. It is clear that any supergauge invariant constructed out of $V^{++}$ and
$V^{--}$ cannot involve $v_0$. So we can from the very beginning set $v_0 = 0 $ in \p{WZ2}

To construct the superfield invariants, we need to know the precise form of the
non-analytic potential corresponding
to the choice \p{WZ2}. It can be found by solving eq. \p{basicVV}. This is
straightforward, though a little bit boring.
One substitutes  the WZ expression \p{WZ2} for $V^{++}$ and the most general
$\theta $ and $u$
expansion of the non-analytic harmonic superfield $V^{--}$ into \p{basicVV}.
Then one uses the explicit expressions
\p{D2} for $D^{\pm\pm}$ and solves \p{basicVV} in components by equating to zero
the coefficients
in front of independent $\theta $ and $u$ monomials. The answer for $V^{--}$ in
the considered bosonic
approximation is as follows
\be
V^{--}(t, \theta^\pm, \bar\theta^\pm) &=& v^{--}
+ \theta^{+\alpha}v^{-3}_\alpha + \bar\theta^+_\alpha \bar v^{-3\alpha} + (\theta^+)^2v^{-4} +
(\bar\theta^+)^2\bar v^{-4} \nn
&& +\, \theta^{+\alpha}\bar\theta^{+\beta}C^{-4}_{(\alpha\beta)} \lb{V--}
\ee
where
\be
&& v^{--} = \sqrt{2}(\theta^-)^2\bar\phi + \sqrt{2}(\bar\theta^-)^2 \phi
+ 2\theta^{-\alpha}\bar\theta^{-\beta}v_{(\alpha\beta)}\,, \; \nn
&& v^{-3}_{\alpha} = -2i (\bar\theta^-)^2\theta^{-\beta}\dot v_{\alpha\beta}
+ 2\sqrt{2}i(\theta^-)^2 \bar\theta^-_\alpha \dot{\bar\phi}\,, \nn
&& \bar v^{-3}_\alpha = 2i (\theta^-)^2\bar\theta^{-\beta}\dot v_{(\alpha\beta)}
- 2\sqrt{2}i(\bar\theta^-)^2 \theta^-_\alpha \dot{\phi}\,, \nn
&& v^{-4} = - \sqrt{2}(\theta^-)^2(\bar\theta^-)^2 \ddot{\bar\phi}\,, \; \bar v^{-4} =
-\sqrt{2}(\theta^-)^2(\bar\theta^-)^2 \ddot{\phi}\,, \nn
&& C^{-4}_{(\alpha\beta)} = - 2(\theta^-)^2(\bar\theta^-)^2\ddot{v}_{(\alpha\beta)}\,. \lb{V--2}
\ee
Now it is straightforward to find $\bar W$ and $W$ defined by eqs. \p{Wdef}
\be
&& \bar W = \sqrt{2}\bar\phi +2\sqrt{2}i(\theta^+\bar\theta^-)\dot{\bar\phi}
- 2i \bar\theta^{+\alpha}\bar\theta^{-\beta}\dot{v}_{(\alpha\beta)} \nn
&& -\,\sqrt{2}\,(\bar\theta^-)^2\left[(\theta^+)^2 \ddot{\bar{\phi}}+ (\bar\theta^+)^2\ddot{\phi}
+ \sqrt{2}(\theta^{+\alpha} \bar\theta^{+\beta})\ddot{v}_{(\alpha\beta)}\right], \lb{W} \\
&& W = \sqrt{2}\phi +2\sqrt{2}i(\theta^-\bar\theta^+)\dot{\phi}
- 2i \theta^{+\alpha}\theta^{-\beta}\dot{v}_{(\alpha\beta)} \nn
&& -\,\sqrt{2}\,(\theta^-)^2\left[(\theta^+)^2 \ddot{\bar{\phi}}+ (\bar\theta^+)^2\ddot{\phi}
+ \sqrt{2}(\theta^{+\alpha} \bar\theta^{+\beta})\ddot{v}_{(\alpha\beta)}\right]. \lb{barW}
\ee

The specificity of the considered $(0+1)$ dimensional case is the appearance of one
additional gauge invariant object
\be
W_{\alpha\beta} = {1\over 2}\bar D^+_{(\alpha}D^+_{\beta)}V^{--}. \lb{Wnew}
\ee
Indeed, using \p{Dcomm}, \p{DD} and the analyticity of the gauge parameter
$\Lambda(\zeta, u)$ it is easy to check invariance
of \p{Wnew} under the gauge transformation \p{gauge42}. Like in the case of $3r$
sigma models, this invariant is none other
than the spatial part of the original vector superconnection. From the definition
\p{Wnew}
it also follows that $W_{\alpha\beta}$ does not depend on harmonics
\be
D^{\pm\pm}W_{\alpha\beta} = 0\,, \lb{indep2}
\ee
and obeys the constraints
\be
D^{\pm(\alpha}W^{\beta\gamma)} = \bar D^{\pm(\alpha} W^{\beta\gamma)} = 0\,.
\ee
In terms of the bosonic components in WZ gauge it is written as
\be
&& W_{(\alpha\beta)} = v_{(\alpha\beta)}
+ 2i\theta^{+\delta}\bar\theta^-_{(\alpha}\dot{v}_{\beta)\delta} +
2i\bar\theta^{+\delta}\theta^-_{(\alpha} \dot{v}_{\beta)\delta}
+  2\sqrt{2}i\theta^+_{(\alpha}\theta^-_{\beta)}\dot{\bar\phi} \nn
&& +\, 2\sqrt{2}i\bar\theta^+_{(\alpha}\bar\theta^-_{\beta)}\dot{\phi}
+ 2\sqrt{2}(\theta^+)^2\bar\theta^-_{(\alpha}\theta^-_{\beta)}\ddot{\bar\phi}
+2\sqrt{2}(\bar\theta^+)^2\bar\theta^-_{(\alpha}\theta^-_{\beta)}\ddot{\phi} \nn
&& +\, 4\bar\theta^-_{(\alpha}\theta^-_{\beta)}
\theta^{+\delta}\bar\theta^{+\sigma}\ddot{v}_{(\delta\sigma)}\,. \lb{Wab}
\ee

Before proceeding further, let us point out that the above ${\cal N}=4$ harmonic
formalism in $d=(0+1)$ actually reveals a hidden
$USp(4) \sim SO(5)$ covariance, besides the manifest covariance under $SU(2)_R$ acting
on the indices $i,k$ of harmonics
and fields in the harmonic expansions and $SU(2)$ realized on the indices $\alpha,
\beta$ (the latter is a remnant of four-dimensional $SL(2,C)$).
This $USp(4) \sim SO(5)$ contains as subgroups both the
latter $SU(2)$ and the $U(1)$ R-symmetry group which acts as a phase
transformation on the $\theta$ s and their conjugate (also on $W$ and $\bar W$ which
have the same $U(1)$ charges as $(\bar D)^2$
and $(D)^2$). The remaining coset $SO(5)/U(2)$
transformations act e.g. on spinor derivatives as
\be
\delta D^\pm_\alpha = \lambda_{\alpha\beta} \bar D^{\pm\beta}\,, \quad \delta \bar
D^{\pm}_\alpha
= -\bar{\lambda}_{\alpha\beta} D^{\pm\beta}\,. \lb{SO5}
\ee
Here $\lambda_{\alpha\beta} = \lambda_{\beta\alpha}$ is a complex triplet
comprising 6 real group parameters.
The harmonic potentials $V^{\pm\pm}$ are $SO(5)$
singlets, while the superfield strengths $W, \bar W, W_{\alpha\beta}$ constitute an
$SO(5)$ vector, as it follows from their
definition \p{Wdef} and \p{Wnew} :
\be
\delta \bar W = \lambda^{\alpha\beta}W_{\alpha\beta}\,,
\, \delta W = \bar\lambda^{\alpha\beta}W_{\alpha\beta}\,,\,
\delta W_{\alpha\beta} = -\lambda_{\alpha\beta}W - \bar\lambda_{\alpha\beta}\bar W.
\ee

One can make  the $SO(5)$ symmetry manifest introducing real superfields $W_I$ with
the  components
$W_{1,2,3}$  related to $W_{\alpha\beta}$ as in eq.(\ref{spin-vect}) and
$W = W_4 + i W_5$.
The $SO(5)$ covariance will be used later to derive  in the HSS approach the
 metric of symplectic sigma models of the second kind defined on $5r$--dimensional
target spaces.

Let us discuss first the case $r=1$. Our basic statement is that a generic
$(0+1)$ action is written in the same way as   the
 general $4D$ action \p{actSW1}, but the function $F$ may depend, besides $W$ and
$\bar W$, on additional gauge invariants $W_{\alpha\beta}$:
\be
\label{actN4}
  S \ =\ \frac 14 \int dt d^8\theta du V^{++} V^{--} F(\bar W, W, W_{\alpha\beta})\,.
\ee
 Like in the $4D$ case, the requirement of gauge invariance
of \p{actN4}
imposes stringent constraints on the function $F(\bar W, W, W_{\alpha\beta})\,$. To
derive them,
one can proceed in the same way as in deriving eq. \p{Zupeq}. Using the relation
between
the integration measures in the full harmonic superspace and its analytic
subspace
$$
dt\, du\, d^8\theta = dt\, du\, (D^-)^4(D^+)^4\; \equiv \; du\, d\zeta^{-4}\,
(D^+)^4\,,
$$
as well as the analyticity of the gauge parameter $\Lambda$ and the relations
\p{+-com}, \p{++-com},
one can cast
the gauge variation of \p{actN4} into the form
$$
\delta S = - \frac 12 \int dt\, du \,d\zeta^{-4}\,\Lambda\,(D^+)^4\left[D^{--}V^{++} F(W,
\bar W, W_{\alpha\beta})\right],
$$
whence the general condition of gauge invariance of \p{actN4} is obtained as
\be
(D^+)^4\left[D^{--}V^{++} F(W, \bar W, W_{\alpha\beta})\right] = 0\,. \lb{GI01}
\ee

When acting by the spinor derivatives from $(D^+)^4$ on the expression within square
brackets,
one obtains a few terms which include derivatives of $F$ with respect to its
superfield arguments,
starting with the second-order ones, and some independent superfield projections of
$V^{++}$
resulting from the action of spinor derivatives on $D^{--}V^{++}$. These independent
contributions
should disappear on their own right. Fortunately, a careful analysis shows that the
common condition of
vanishing of all these terms is the following constraint which is the $(0+1)$ analog
of \p{Zupeq} \cite{Zupn}
\be
\left(\frac{\partial^2}{\partial W^{\alpha\beta}\partial W_{\alpha\beta}} +  2
\frac{\partial^2}{\partial W \partial \bar W}\right) F = 0\,. \lb{Zupeq2}
\ee

The simplest way to deduce \p{Zupeq2} is to consider in \p{GI01} one of the terms which
are produced when
$(D^+)^4$ as a whole is pulled out through $D^{--}V^{++}$ and hits the function $F$.
This specific term contains
only bosonic superfields and is a collection of the terms which are obtained either
when $(D^+)^2$, $(\bar D^+)^2$
hit, respectively, $W$, $\bar W$ or $D^+_\alpha \bar D^+_\beta$ hits
$W_{\gamma\lambda}$. Using the relations
\be
(D^+)^2W = (\bar D^+)^2 \bar W \equiv G^{+2}\,, \; D^+_\alpha \bar D^+_\beta
W_{\gamma\lambda} =
{1\over 4}\,(\epsilon_{\alpha\gamma}\epsilon_{\beta\lambda}
+\epsilon_{\alpha\lambda}\epsilon_{\beta\gamma}) G^{+2}
\ee
which follow from the explicit expressions \p{Wdef} and \p{Wnew}, one can show that
this term in \p{GI01}
is represented as
\be
\propto \; D^{--}V^{++}\, (G^{+2})^2 \left(\frac{\partial^2}{\partial
W^{\alpha\beta}\partial W_{\alpha\beta}} +  2
\frac{\partial^2}{\partial W \partial \bar W}\right) F\,,\lb{specterm}
\ee
whence \p{Zupeq2} is obtained as the condition of its vanishing. All other terms
 involve the same structure or its derivatives and  vanish together with
\p{specterm}.

Eq. \p{Zupeq2} is of course nothing but a familiar 5-dimensional Laplace equation
like in \p{00}. Note that in the ${\cal N} = 2$ approach of sect. 2, the harmonicity constraint
is imposed on the prepotential ${\cal K}$,
while in the ${\cal N} = 4$ approach of the present section it is imposed on the function
$F(W_I)$.
It remains to learn how these two functions are related.

The most explicit  way to do this is to express the action \p{actN4} in components.
The computations
in the general case become a little bit sophisticated  since \p{actN4}
involves the integral $\int d^8 \theta$, which cannot be reduced to
 an
integral over chiral superspace like it was the case in four dimensions
( see \p{actSW}): in contrast to $W$ and $\bar W$,     $W_{\alpha\beta}$ is not chiral.
Nevertheless, the calculations are feasible.
After substituting the explicit expressions \p{WZ2}, \p{V--}, \p{V--2}, \p{W}, \p{barW}
and \p{Wab} into \p{actN4}, doing there the $u$- and $\theta$-integrals,
and going over to $SO(5)$ vector notation, one obtains
\be
{\cal L}_{\rm kin} = {1\over 2} \dot{v}_I^2
\left(1 + {3\over 2} \hat D + {1\over 2} \hat D^2\right) F(v_I) \equiv {1\over 2}\, \dot{v}_I^2\, h(v_I)
\lb{result1}
\ee
where
\be
v_I \equiv \left(v_k, v_4, v_5\right), \quad
v_{\alpha\beta} = i\epsilon_{\alpha\gamma}(\sigma_k)^\gamma_\beta v_k\,, \quad \phi =\frac{v_4 +iv_5}{\sqrt{2}}\,,
\ee
and
\be
\hat D \ \equiv \ v_I \frac {\partial}{\partial v_I}
\lb{Oper}
\ee
is the operator of the target space scale transformation. Now, the commutator of
$\hat D$ and the $5D$ Laplacian $\Delta^{(5)}$ gives again  $\Delta^{(5)}$, and this
means that
the function
    \be
   \label{hf}
   h (v_I) = \left(1 + {3\over 2} \hat D +
   {1\over 2}\hat D^2 \right)F
    \ee
is harmonic provided $F$ is harmonic.
It can
be identified, up to a numerical coefficient, with the harmonic function $h$
introduced in  \p{metr5}. The simplest example is $SO(5)$ invariant metric \cite{DE}
for which
\be
h (v_I) = F(v_I) = c_0 +  c_1\,(v_Iv_I){}^{-3/2} \lb{so5inv}
\ee
where $c_0, c_1$ are some constants.

The prepotential ${\cal K}$ can be restored by solving the 3-dim Laplace equation
  \be
  2 \frac {\partial^2 {\cal K}}{\partial v_k^2} \ =\ h
   \ee
   (cf. eq. (2.10) in ref.\cite{N2}).

There is another way to arrive at the same result (\ref{hf}), which exploits
 $SO(5)$ covariance of the ${\cal N}=4, (0+1)$ formalism.
Let us start from the $(0+1)$ reduction of the
Seiberg--Witten action (\ref{actSW}) having no dependence on the superfield $W_k$,
i.e. with $F(W, \bar W) = f(W) + \bar f(\bar W)$. The metric
in the space of scalar fields $\phi$, $\bar\phi$ is just
  \be
  \label{hfi}
h(\bar\phi, \phi) &=& \frac 12 \frac {\partial^2}{\partial \phi^2} \left[
\phi^2 f(\phi) \right] + {\rm c.c.} \nonumber \\
&=& \left[1 - 2 \left(\phi \frac {\partial}{\partial \phi } +   \bar\phi \frac
{\partial}{\partial \bar\phi } \right)  +
\frac 12 \left(\phi^2 \frac {\partial}{\partial \phi^2 } +   \bar\phi^2 \frac
{\partial^2}{\partial \bar\phi^2 } \right) \right] F\ .
   \ee
   Bearing in mind that $\phi = (v_4 + i v_5)/\sqrt{2}$, one can rewrite it as
   \be
    \label{hA45}
    h \ =\ \left[1 + \frac 32 \left(v_4 \frac {\partial }{\partial v_4} +
    v_5\frac {\partial }{\partial v_5}\right) + \frac 12 \left(v_4 \frac
{\partial }{\partial v_4}
    + v_5\frac {\partial }{\partial v_5}\right) ^2   \right] F\ .
  \ee
Allowing now for the dependence on $v_k$, $F(\phi,\bar \phi) \rightarrow F(\phi,\bar
\phi, v_k)$,  and requiring
the $SO(5)$ covariance, we arrive at eq.(\ref{hf}).
To avoid a misunderstanding, we should point out that the general metric is not
obliged to have $SO(5)$ as an isometry. But the
whole breaking of $SO(5)$ is encoded in the structure of the function $F$, while
the relation between this function and metric is
$SO(5)$ covariant by construction. Indeed, all steps leading from \p{actN4} to the
kinetic sigma model term \p{result1} preserve $SO(5)$ covariance.

The action (\ref{actN4}) can be easily generalized to the case of several multiplets.
We write
   \be
\label{actN4many}
  S \ =\  \int dt d^8\theta du \ V^{++}_A V^{--}_B F^{AB}(\bar W^C, W^C, W_{k}^C)
  \ee
where $F^{AB}$ can be assumed to be symmetric under $A \leftrightarrow B$. To show
this, one should express $V^{--}_B$ via
$V^{++}_B$ to derive (cf. eq.(7.26) in ref.\cite{kniga})
  \be
\label{actsymAB}
  S \ =\  \int dt d^8\theta du_1 du_2 \frac  {V^{++}_A(t,\theta, u_1)
V^{++}_B (t,\theta, u_2)}{(u^+_1 u^+_2)^2}
 F^{AB}( W_{I}^C).
  \ee
Using the fact that $W^C_I$  do not depend on harmonics, we see that the
antisymmetric part of $F^{AB}$ does not contribute.

The action (\ref{actN4many}) is gauge invariant provided
     \be
\label{harmf AB}
\frac {\partial^2 F^{AB}}{\partial W_I^{C} \partial W_I^{D}}\ =\
 \frac {\partial^2 F^{AB}}{\partial W_I^{[C} \partial W_J^{D]}}
\ =\ 0\ .
    \ee
These conditions can be proved like in the earlier considered case
of one multiplet.

To derive the metric, we proceed in the same way as in the one-flavor case.
Suppressing the dependence on $W_k^C$, we arrive at
  \be
  \label{actSWAB}
S = \int dt d^4\theta\ W^A W^B \tilde{f}^{AB}(W^C) \ +\ {\rm c.c.}
 \ee
 (the Seiberg--Witten effective action for a group of rank $r$ has such a form).
 Deriving the metric $h^{AB} (\bar\phi^C, \phi^C)$ and restoring the $SO(5)$ covariant
dependence on $v_k^C$, we arrive at the result
   \be
   h^{AB} &=& F^{AB} + \frac 34 (R^{CB} F^{CA} + R^{CA} F^{CB} ) \nn
   && +\,
   \frac 14 (R^{CA} R^{DB} + R^{DB} R^{CA} ) F^{CD} \lb{ResultN}
   \ee
   where $R^{CA} = v_I^C \partial_I^A$. It is straightforward to check that this
$h^{AB}$ satisfies
   the same conditions \p{harmf AB} as $F^{AB}$. This is in the full agreement with
the conditions
   \p{harm5r} derived in the ${\cal N} = 2$ superfield approach.

To summarize, the multiflavor analog of the sigma model Lagrangian \p{result1} is
\be
{\cal L}^{mf}_{kin} = {1\over 2}\,\dot{v}^A_I\dot{v}^B_I\, h^{AB}(v)\,. \lb{ResultN2}
\ee
The symmetric matrix function $h^{AB}(v)$ satisfies eqs. \p{harmf AB} (with $W^C_I \rightarrow v^C_I$)
and is arbitrary otherwise.
Its precise relation to the function ${\cal K}$ of the ${\cal N}=2$ superfield formalism remains
to be clarified.

\section{Conclusions}
In this paper we studied the general off-shell actions of ${\cal N}=4, (0+1)$
supersymmetric sigma models associated with the ${\cal N}=4$ multiplets $({\bf 5, 8, 3})$
(symplectic sigma models of the second type), both for a single supermultiplet
and for several such multiplets. We performed our analysis in
the ${\cal N}=2$ superspace where only half of the underlying ${\cal N}=4$ supersymmetry
is manifest and in the harmonic ${\cal N}=4$ superspace which makes manifest the entire
supersymmetry. In the first formalism the multiplet in question is represented by
the ${\cal N}=2$ superfields $V_I \propto (V_k, \Phi, \bar\Phi)$ with the off-shell contents
$({\bf 3, 4, 1})$ and $({\bf 2, 4, 2})$, respectively. The most general ${\cal N}=2$
action is given by \p{act5many} where the Lagrangian ${\cal K}(V^A_I)$ obeys the constraints \p{harm5r}.
In the harmonic ${\cal N}=4$ formalism the same multiplet is presented by the
harmonic analytic gauge superfield $V^{++}$,  from which in $(0+1)$
dimension one can construct the covariant superfield strengths $(W_{(\alpha\beta)}, W, \bar W) \equiv W_I$
with 5 physical bosonic fields as their first components. However, the most general action cannot be
written as a superspace integral of some gauge invariant Lagrangian composed out of these
covariant strengths; it is inevitably of the Chern-Simons type and is gauge invariant
only modulo a shift of the integrand by a total harmonic derivative. It is given by the
expression \p{actN4many}, with the function $F^{AB}(W_I^C)$ obeying the constraints \p{harmf AB}. We explicitly
computed the target bosonic metric in terms of this function in the one-flavor case (eq. \p{result1})
and restored the metric in the multiflavor case by using the $SO(5)$ covariance of the
harmonic formalism (eqs. \p{ResultN}, \p{ResultN2}).
\footnote{ Whereas the ${\cal N}=4$ actions  \p{actN4}, \p{actN4many} were proposed earlier in \cite{Zupn},
the explicit calculation of the target bosonic metrics both for the one-flavor and
multiflavor cases is our new result. Also, the general constraints \p{harmf AB}
were not given in \cite{Zupn}.}

It would be interesting to understand the target bosonic geometry associated with the actions \p{act5many}
and \p{actN4many} along the line of e.g. \cite{PCo}, \cite{Hull} and to reveal the geometric meaning of the
constraints \p{harm5r}, \p{harmf AB}. It is also desirable to realize
how these models are inscribed into the general ``Gauge Theories/Strings'' correspondence.
As mentioned in Introduction, for the symplectic sigma models of the first type such an interpretation is quite likely
(they emerge both in the context of the low-energy description of ${\cal N}=1, d =4$ gauge theories
and in the black holes stuff). Meanwhile, so far only the first, gauge theory ``face'' of the symplectic models
of the second type has been clarified \cite{N2}. It is still unclear what could be the string theory ``face''
of these sigma models. Some hints are given in \cite{DE} and \cite{Tow}. In particular, it was argued in
\cite{Tow} that the superconformal versions of such models can describe the near-horizon AdS$_2\times S^4$ geometry
of a D5-brane in an orthogonal D3-brane background. Note that the superconformal action for one-flavor
symplectic model of the second type was constructed in the framework of the ${\cal N}=2$ superfield
formalism in a recent preprint \cite{BIKL} (its bosonic sector corresponds to the choice
$c_0 = 0, c_1 \neq 0$ in \p{so5inv}).

As a final remark we note that there exists an alternative
${\cal N}=2$ description of the same ${\cal N}=4$ multiplet $({\bf 5, 8, 3})$ in terms of the
multiplets with the off-shell contents $({\bf 1, 4, 3})$ and $({\bf 4, 4, 0})$ \cite{BIKL}
(the latter multiplet involves no auxiliary fields at all). It is of obvious interest
to see what the generic action of the considered class of symplectic sigma models
looks like in this alternative formulation.

\section*{Acknowledgements}
We thank B.M. Zupnik for discussions. The work of E.I. was partially supported by INTAS grant 00-00254,
RFBR-DFG grant 02-02-04002, grant DFG 436 RUS 113/669, RFBR grant
03-02-17440, a grant of the IN2P3-JINR program and a grant
of the Heisenberg-Landau program. He thanks SUBATECH of the University of Nantes  for the kind hospitality
at the early stage of this study. A.S. acknowledges the warm hospitality in BLTP in Dubna.

\end{document}